\documentclass[aps, secnumarabic, nobibnotes, twocolumn, superscriptaddress]{revtex4-2}

\usepackage{amsfonts}
\usepackage{mathrsfs}
\usepackage{amsmath}
\usepackage{color}
\usepackage{natbib}
\usepackage{graphicx}
\usepackage{bm}
\usepackage{amssymb}
\usepackage{xspace}
\usepackage{epstopdf}
\usepackage{dcolumn}
\usepackage{longtable}
\usepackage{multirow}
\usepackage[title,titletoc]{appendix}
\usepackage{graphicx}
\usepackage{float}
\usepackage[colorlinks=true, letterpaper=true, pdfstartview=FitV, linkcolor=blue, citecolor=blue, urlcolor=blue]{hyperref}

\makeatletter

\newcommand{\Rmnum}[1]{\expandafter\@slowromancap\romannumeral #1@}
\makeatother
\begin{document}

\title{Circularly polarized light irradiated ferromagnetic MnBi$_2$Te$_4$: the long-sought ideal Weyl semimetal}

\author{Shuai Fan}
\thanks{These authors contributed equally to this work.}
\affiliation{%
Institute for Structure and Function $\&$ Department of Physics $\&$ Chongqing Key Laboratory for Strongly Coupled Physics, Chongqing University, Chongqing 400044, People's Republic of China
}%

\author{Shengpu Huang}
\thanks{These authors contributed equally to this work.}
 \affiliation{%
Institute for Structure and Function $\&$ Department of Physics $\&$ Chongqing Key Laboratory for Strongly Coupled Physics, Chongqing University, Chongqing 400044, People's Republic of China
}%

\author{Zhuo Chen}
\thanks{These authors contributed equally to this work.}
 \affiliation{%
Institute for Structure and Function $\&$ Department of Physics $\&$ Chongqing Key Laboratory for Strongly Coupled Physics, Chongqing University, Chongqing 400044, People's Republic of China
}%
\author{Fangyang Zhan}
 \affiliation{%
Institute for Structure and Function $\&$ Department of Physics $\&$ Chongqing Key Laboratory for Strongly Coupled Physics, Chongqing University, Chongqing 400044, People's Republic of China
}%
\author{Xian-Yong Ding}
 \affiliation{%
Institute for Structure and Function $\&$ Department of Physics $\&$ Chongqing Key Laboratory for Strongly Coupled Physics, Chongqing University, Chongqing 400044, People's Republic of China
}%
\author{Da-Shuai Ma}%
\email{madason.xin@gmail.com}
\affiliation{%
Institute for Structure and Function $\&$ Department of Physics $\&$ Chongqing Key Laboratory for Strongly Coupled Physics, Chongqing University, Chongqing 400044, People's Republic of China
}%
\affiliation{%
Center of Quantum materials and devices, Chongqing University, Chongqing 400044, People's Republic of China
}%
\author{Rui Wang}%
 \email{rcwang@cqu.edu.cn}
 \affiliation{%
Institute for Structure and Function $\&$ Department of Physics $\&$ Chongqing Key Laboratory for Strongly Coupled Physics, Chongqing University, Chongqing 400044, People's Republic of China
 }%
 \affiliation{%
Center of Quantum materials and devices, Chongqing University, Chongqing 400044, People's Republic of China
 }%


\begin{abstract}
The interaction between light and non-trivial energy band topology allows for the precise manipulation of topological quantum states, which has attracted intensive interest in condensed matter physics.
In this work, using first-principles calculations, we studied the topological transition of ferromagnetic (FM) MnBi$_2$Te$_4$ upon irradiation with circularly polarized light (CPL). 
We revealed that the MnBi$_2$Te$_4$ can be driven from an FM insulator to a Weyl semimetal with a minimum number of Weyl points, i.e., two Weyl points in systems without time-reversal symmetry.
More importantly, in FM MnBi$_2$Te$_4$ with out-of-plane easy magnetization axis, we found that the band dispersion of the WP evolves from Type-II to Type-III and finally to Type-I when the light intensity increases.
Moreover, we show that the profile of the characteristic Fermi arc of Weyl semimetal phase is sensitive to changes in light intensity, which enables efficient manipulation of the Fermi arc length of FM MnBi$_2$Te$_4$ in experiments.
In addition, for FM MnBi$_2$Te$_4$ with in-plane easy magnetization axis, the system becomes a type I Weyl semimetal under CPL irradiation.
With controllable band dispersion, length of Fermi arc, and minimum number of WPs, our results indicate that CPL-irradiated FM MnBi$_2$Te$_4$ is an ideal platform to study novel transport phenomena in Weyl semimetals with distinct band dispersion.
\end{abstract}

\maketitle

\section{\label{sec:level1}INTRODUCTION}

Topological semimetals, a class of phases with energy band crossings near the Fermi level, have sparked growing interest due to the extension of topological classification and promising realizations of elementary particles~\cite{RevModPhys.82.3045,RevModPhys.83.1057,RevModPhys.90.015001,TSfFP,burkov_topological_2016}. 
According to band degeneracy, topological semimetals can be distinguished into Dirac semimetals~\cite{Dirac4,Dirac5,Dirac6,Dirac7}, Weyl semimetals (WSM)~\cite{Weyl1,Weyl2,Weyl3,Weyl4}, nodal-line semimetals~\cite{line1,line2,line3}, as well as beyond~\cite{beyond1,beyond2,beyond3,beyond4,beyond5}. 
Among these topologically nontrivial materials, WSM exhibits novel transport phenomena and is one of the typical representatives.
WSMs are featured by Weyl points (WPs) consisting of energy bands with linear dispersion in momentum space, and fermionic excitation near these WPs is depicted by two-component Weyl equation~\cite{TMWS}. 
It is realized that WPs could be regarded as sources of quantized Berry flux. 
Dependent the source and drain of Berry flux, the charges of the WPs (i.e. chirality) are defined to be $+1$ and $-1$, respectively. 
Presently, based on the manifold of the Fermi surface, WSMs can be classified into three types~\cite{type24,type25}. 
Type-I Weyl semimetal exhibits standard WPs with the Fermi surface shrinks to a point near the band crossing~\cite{Weyl3}. 
Type-II Weyl semimetal has a significant tilt behaviour and is characterized by the coexistence of electron and hole pockets connected by WPs.
Between type-I and type-II Weyl semimetal, there is a critical point featured by the appearance of flat band along specific direction in momentum space, termed type-III Weyl semimetal.
The Fermi surface of a type-III Weyl semimetal is a single line, which results in highly anisotropic effective masses and finite density of states~\cite{type28}. 
Besides the distinct Fermi surface, it is theoretically predicated that the amplitude of the tilt behaviour of WPs also lead to the emergence of intricate transport phenomena, such as modified Klein tunneling and unconventional Landau levels~\cite{PhysRevLett.116.236401,PhysRevLett.117.077202,PhysRevB.97.235113,yuan2018chiral,PhysRevB.107.085146}.
To achieve experimental verification of those above-mentioned theoretical predictions and obtain a clearer understanding of Weyl semimetal phase, it remains an open question how to select a natural material candidate that can achieve a minimum number of WPs whose band dispersion could be designed as desired by experimental means.

Recently, Floquet engineering (i.e., periodic driving of polarized light) has played a vital role in artificial manipulation of topological quantum states, leading to a series of breakthroughs in non-equilibrium phases~\cite{type26,flo27,flo33,floreal1,PhysRevB.98.235159,floreal2,flo28,flo32,zhan2023floquet,PhysRevB.107.L121407,flo34,flo35,flo36}. 
There are two essential reasons for the feasibility of manipulating topological states through irradiation of light: (I) the symmetry breaking by incident light, and (II) the $k$-dependent vector-potential of light would modify the band structure and leads to possible band inversions.
For example, due to the fact that circularly polarized light (CPL) breaks time-reversal symmetry (TRS), it is theoretically predicted and experimentally verified that the irradiation of CPL can introduce a certain Dirac mass term to the Dirac cones of the surface states of topological insulators, thereby resulting in Chern insulators~\cite{floreal1,floreal2}. 
Moreover, CPL has been proposed as an efficient means to drive nodal line semimetals and Dirac semimetals into Weyl semimetals~\cite{PhysRevLett.117.087402,PhysRevB.105.L081102,PhysRevB.108.205139}.
While significant progress has been made in this field, the manipulation of band dispersion in Weyl semimetals via Floquet engineering in natural materials remains largely unexplored.

In this work,  utilizing first-principles calculations, we meticulously tracked the band gap evolution of FM MnBi$_2$Te$_4$ with out-of-plane and in-plane easy magnetization axis under the irradiation of CPL, respectively.
We found that both FM phases of MnBi$_2$Te$_4$ can be driven from an FM insulator to an ideal Weyl semimetal with the minimum number of WPs (two Weyl points for system without TRS).
For FM MnBi$_2$Te$_4$ with the out-of-plane easy magnetization axis, we observed that increasing the intensity of circularly polarized light (CPL) induces a sequential transformation in the band dispersion of Weyl points (WPs). Specifically, this progression transitions from a type-II dispersion to type-III, culminating in a Type-I dispersion.
The characteristic Fermi arcs and surface states of the Weyl semimetal phase under distinct light intensities were investigated. 
We found that the profile of the characteristic Fermi arcs on (010) surface of FM MnBi$_2$Te$_4$ irradiated with CPL exhibits sensitivity to variations in light intensity, enabling effective control of the Fermi arc length through the incident light.
At the end of this work, we reoriented the magnetic orientation of MnBi$_2$Te$_4$ to the in-plane axis.
We found that, in this case, the system could also transition to a Weyl semimetal with two WPs as the light intensity increases.
The WPs found in CPL-irradiated FM MnBi$_2$Te$_4$ with in-plane easy magnetization axis maintain type-I band dispersion until they annihilation. 

\section{computational methods\protect}
First-principles calculations were performed by Vienna $Ab$ $initio$ Simulation Package~\cite{xx56,xx57} in the framework of density functional theory (DFT)~\cite{xx58,xx59}. 
The electron-ion interaction was treated by projector-augmented-wave potentials~\cite{xx61}. 
The generalized gradient approximation (GGA) with Perdew-Burke-Ernzerhof (PBE) formalism was employed to describe the exchange-correlation functional~\cite{xx60}. 
The experimental lattice constants are adopted~\cite{otrokov_prediction_2019}.
The energy cutoff of the plane wave basis was set to 350 eV, and a k-mesh with a 9 × 9 × 5 Monkhorst–Pack grid~\cite{xx62} in momentum space was used to determine the band structure.
Spin-orbit coupling (SOC) effects were taken into account in the self-consistent calculations. 
To deal with the correlation effects of 3$d$ electrons in Mn atoms, we employed the GGA+$U$ method~\cite{xx63} and set the $U=4$ eV. 
In the process of structural relaxation, the forces on each atom were relaxed to be less than 0.02 eV/{\AA}. 
The topological properties were revealed by constructing a Wannier-function-based tight-binding (WFTB) Hamiltonian based on maximally localized Wannier functions methods combining DFT calculations~\cite{xx66,xx67,xx68}. 
Employing the WFTB Hamiltonian, we derived a time-dependent Hamiltonian through the Peierls substitution. 
The topological surface states and Fermi arcs calculated using the WANNIERTOOLS package~\cite{xx70,xx71}.

\section{results and discussion\protect}
\begin{figure}[t]
\includegraphics[width=\linewidth]{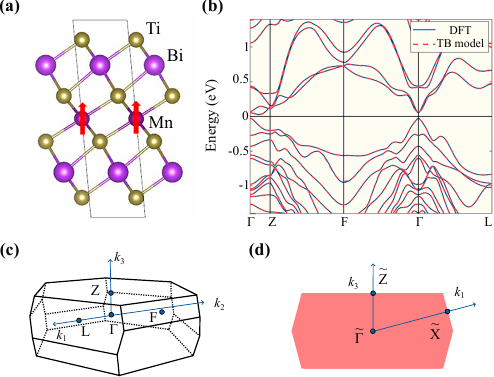}
\caption{
(a) Crystal and magnetic structures of FM-$z$ MnBi$_2$Te$_4$. The magnetic moments of the Mn atoms are directed along the red arrows.. 
(b) Electronic structure of FM-$z$ MnBi$_2$Te$_4$ obtained from DFT calculations (solid blue line) and Wannier-function-based tight-binding model (dashed red line).
(c) Brillouin zone for the conventional cell shown in (a) with high-symmetry points marked. (d) Surface BZ for a side surface, i.e., (010) surface.
}
\label{fig1}
\end{figure}

MnBi$_2$Te$_4$ belongs to a rhombohedral structure with the space group $R\overline{3}m$  (No.~166)~\cite{lee_crystal_2013}. 
As depicted in Fig.~\ref{fig1}(a), the MnBi$_2$Te$_4$ single crystal is composed of Te-Bi-Te-Mn-Te-Bi-Te septuple layers stacking along the $c$ axis.
Two adjacent septuple layers are coupled to each other via van der Waals forces.
The experimental lattice parameters of the unit cell are $\left|\boldsymbol{a}\right|=\left|\boldsymbol{b}\right|=4.33$ $\mathrm{\AA}$ and $\left|\boldsymbol{c}\right|=40.91$ $\mathrm{\AA}$.
In the calculation of the electronic properties for FM MnBi$_2$Te$_4$, we select the lattice vectors as $\boldsymbol{a}'=\boldsymbol{a}$, $\boldsymbol{b}'=\boldsymbol{b}$, and $\boldsymbol{c}'=\left[\boldsymbol{c}-\left(\boldsymbol{a}+\boldsymbol{b}\right)\right]/3$.
It is reported that, due to the Anderson superexchange, the magnetic ground state of bulk MnBi$_2$Te$_4$ is A-type AFM phase with the out-of-plane easy magnetization axis.
Although the total energy of the FM phases is approximately 0.046 eV higher than that of the A-type AFM phase, the field-induced FM phase of MnBi$_2$Te$_4$ is reported to be experimentally achievable~\cite{deng2020quantum}.  
Here, we mainly consider FM-$z$ phase of MnBi$_2$Te$_4$ with the out-of-plane easy magnetization axis as shown in Fig.~\ref{fig1}(a). 
We examine the topological evolution of the band structure in the FM phase MnBi$_2$Te$_4$ under periodic CPL irradiation.
Additionally, other ferromagnetic (FM) phases of MnBi$_2$Te$_4$, such as the FM-$y$ phase with an in-plane easy magnetization axis aligned along the $y$-axis, will be briefly discussed in the concluding section of our work.

The bulk Brillouin zone (BZ) of FM-$z$ MnBi$_2$Te$_4$  is depicted in Fig.~\ref{fig1}(b).
The band structure of the FM-$z$ MnBi$_2$Te$_4$ along high-symmetry lines within the bulk BZ is shown in  Fig.~\ref{fig1}(c).  
It is observed that the system is a FM insulator with an indirect band gap of  $\sim53~$meV.
The smallest local band gap, approximately  $75~$meV,  is located  at the high-symmetry point $\mathrm{\Gamma}$. 
To trace the light-manipulated band topology of FM MnBi$_2$Te$_4$, we employed the maximally localized Wannier functions (MLWF) to construct WFTB from first-principles calculations based on DFT.
According to the fat-band analysis, three $p$ orbitals of Bi and Te atoms and five $d$ orbitals of Mn atoms are selected to initialize the MLWFs. 
The band structure of FM-$z$ MnBi$_2$Te$_4$ obtained from the constructed WFTB are delineated  by the red dashed lines in Fig.~\ref{fig1}(c) and is found to be in agreement with results obtained by DFT calculation. 
Based on this reliable  WFTB, the Peierls substitution is adopted to obtain the time-dependent Hamiltonian, which can capture the electronic properties  of FM MnBi$_2$Te$_4$ when it is irradiated by CPL.

\begin{figure}[t]
\includegraphics[width=\linewidth]{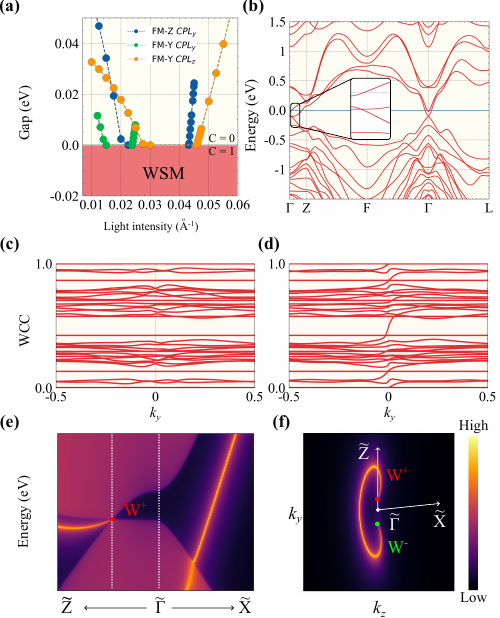}
\caption{
(a) The evolution of band gap of FM MnBi$_2$Te$_4$ with increasing light intensity $eA_{1}/\hbar$. 
In this plot, the results for FM-$z$ MnBi$_2$Te$_4$ irradiated by CPL with polarization in the $x$-$z$ plane are marked in blue, while those  for FM-$y$ MnBi$_2$Te$_4$ irradiated by CPL with polarization in the $x$-$z$ ($x$-$y$) plane are marked in green (orange).
Moreover, the colors of shaded areas denote different Chern number of $k_3=0$ plane. 
(b)Electronic structure of FM-$z$ MnBi$_2$Te$_4$ at light intensity \( eA_{1}/\hbar = 0.0300 \) Å$^{-1}$. The inset is the band structure along $\mathrm{\Gamma}$-$\mathrm{Z}$, where the WP exhibits type-III dispersion.
(c)-(d) The Wannier charge centers of FM-$z$ MnBi$_2$Te$_4$  in the (c) $k_3=\pi$ and (d) $k_3=0$ plane under the irradiation of a CPL with light intensity $eA_{1}/\hbar$ = 0.0300 {\AA}$^{-1}$.
(e) The local density of states of the  (010) surface of FM-$z$ MnBi$_2$Te$_4$. 
(f) The isoenergy band contours of (010) surfaces at $-0.104~$meV relative to the Fermi level.
In (e) and (f), the first BZ of (010) surface is shown in Fig.~\ref{fig1}(d), and the projected position of Weyl points with positive and negative chirality are marked in red and blue,  respectively.
}
\label{fig2}
\end{figure}

Under irradiation with CPL, the time-dependent vector potential is defined  as $\boldsymbol{A}(t)=A_{0}\left[\mathrm{cos}\left(\omega t\right),0,\mathrm{sin}\left(\omega t\right)\right]$, where $A_{0}$ is the amplitude of the CPL, and $\omega$ is the frequency. 
Hence, the intensity of the light is $e A_0/\hbar$.
The incident light irradiates along the $y$ axis and is polarized within the $x$-$z$ plane.
In our calculations, the $\hbar\omega=15$~eV is adopted to avoid interactions between different Floquet sub-bands.
As depicted in Fig.~\ref{fig2}(a), we show the evolution of the band gap of FM-$z$ MnBi$_2$Te$_4$ under light irradiation with varying intensities (i.e., $eA_{0}/\hbar$ = 0.010 {\AA}$^{-1}\mapsto$0.060 {\AA}$^{-1}$).
It was found that light irradiation significantly  changes the energy band profile as the amplitude increases.
When the light intensities exceed \(eA_{0}/\hbar = 0.020\) Å\(^{-1}\), band crossings occur along the high-symmetry line $\mathrm{\Gamma}$-$\mathrm{Z}$, indicating the transformation of FM-$z$ MnBi$_2$Te$_4$ from trivial insulators to semimetals.
Meanwhile, the FM-$z$ MnBi$_2$Te$_4$ becomes a trivial insulator again when the intensity of light exceeds $eA_{0}/\hbar$ = 0.043 {\AA}$^{-1}$. 
Notably, this  critical value of light intensity, corresponding to the electric field strength of 6.459 × 10$^{9}$ V/m or peak intensity of 5.541 × 10$^{12}$ W/cm$^{2}$~\cite{PhysRevB.99.075121,type26,PhysRevLett.120.156406}, can be realized in experiments~\cite{NanoLett79937998,floreal2}.
Through careful searching of local minimum of energy differences between valence and conduction bands, we find that, in the specific range of light intensity (i.e., 0.020 {\AA}$^{-1}\mapsto$0.043 {\AA}$^{-1}$), there are only two WPs pinned on high-symmetry line $\mathrm{\Gamma}$-$\mathrm{Z}$ in the whole bulk BZ.

\begin{figure}[b]
\includegraphics[width=\linewidth]{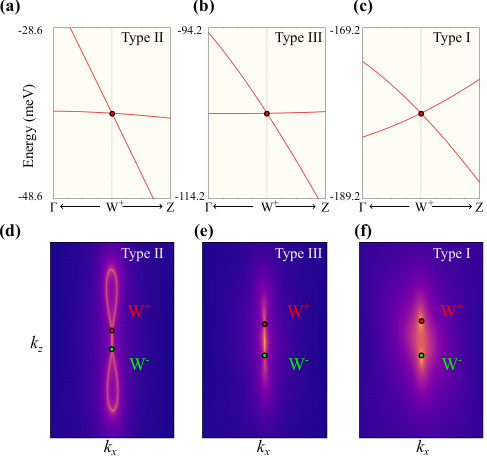}
\caption{(a)-(c) Calculated band structure of FM-$z$ MnBi$_2$Te$_4$ along high-symmetry line $\mathrm{\Gamma}$-$\mathrm{Z}$ with light intensity $eA_{1}/\hbar$ of (a) 0.0225 {\AA}$^{-1}$, (b) 0.0300 {\AA}$^{-1}$, and (c) 0.0375 {\AA}$^{-1}$. 
(d)-(f) Isoenergy contour in the $k_2=0$ plane of the bulk band structure of FM-$z$ MnBi$_2$Te$_4$ with $eA_{1}/\hbar$ of (d) 0.0225 {\AA}$^{-1}$, (e) 0.0300 {\AA}$^{-1}$, and (f) 0.0375 {\AA}$^{-1}$. 
In plots (d)-(f), the energies of the isoenergy contours correspond to those of the respective WPs.}
\label{fig3}
\end{figure}

Let us turn to the FM-$z$ MnBi$_2$Te$_4$ under the irradiation of  CPL with a light intensity of $eA_{0}/\hbar$ = 0.0300 {\AA}$^{-1}$ and detect the band topology of the Weyl semimetal phase.
In this case, as shown in Fig.~\ref{fig2}(b), the two WPs with chirality $\pm 1$  are positioned at $\pm\left(0.000,0.000,0.024\right)$~$\mathrm{\AA^{-1}}$, located at $\sim-0.104~$eV relative to the Fermi level, and are labeled as $W^\pm$.
The energy dispersion of  FM-$z$ MnBi$_2$Te$_4$ with $eA_{0}/\hbar$ = 0.0300 {\AA}$^{-1}$ along $\mathrm{\Gamma}$-$\mathrm{Z}$ is inserted in Fig.~\ref{fig2}(b).
One can observe that the Weyl node $W^+$, exhibiting type-III dispersion, consists of one flat band and one dispersive band.
Due to the remaining inversion symmetry $\mathcal{P}$ under irradiation, the WP $W^-$ keep the same dispersion with $W^+$.
Furthermore, to illustrate the non-trivial band topology of this Weyl semimetal phase, we present the surface state of FM-$z$ MnBi$_2$Te$_4$ with $eA_{0}/\hbar$ = 0.0300 {\AA}$^{-1}$ by using the Floquet WFTB Hamiltonian.
The obtained local density of states (LDOS) projected on the semi-infinite (010) surface is presented in Fig.~\ref{fig2}(e).
It is observed that the characteristic topological Fermi arc terminates at the projected WP $W^+$ and extends along the $\widetilde{\mathrm{\Gamma}}$-$\widetilde{\mathrm{Z}}$ direction.
Moreover, the Fermi arc is found along the $\widetilde{\mathrm{\Gamma}}$-$\widetilde{\mathrm{X}}$ direction.
In order to fully understand the appearance of the Fermi arc, we calculated the evolution of Wannier charge centers~\cite{yu_equivalent_2011}  in the $k_z = 0$ and $\pi$ two-dimensional planes, as illustrated in Figs.~\ref{fig2}(c) and \ref{fig2}(d), respectively.
It is found that there is unavoidable Wilson loop winding in the Wilson loop spectrum of $k{_z} =0$ plane, whereas a gap is present in the spectrum of $k{_z} =\pi$ plane.
Correspondingly, we have the Chern number $\mathcal{C} = 1$ on the $k{_z} = 0 $ plane, and $\mathcal{C} = 0$ on the $k{_z}=\pi$ plane, indicating that when projected onto a plane perpendicular to the $z$ axis, the Fermi arc will intersect the $k_z = 0$ line of the surface BZ an odd number of times.
To confirm this prediction, we further examine isoenergy contours of semi-infinite (010) surface, as shown in  Fig.~\ref{fig2}(f). 
The isoenergy band contours of (010) surfaces at $-0.104~$meV relative to the Fermi level is shown in Fig.~\ref{fig2}(f), where the projected positions of WPs are marked. 
Fig.~\ref{fig2}(f) indicates that there is one Fermi arc connecting the projections of WPs with opposite chirality and intersecting $k_z=0$ line once. 
These two WPs, which are close to the Fermi level, are well separated in momentum space (i.e., 0.060 Å$^{-1}$), and the Fermi arcs are clearly discernible, greatly facilitating experimental observation.

This light-induced WSM, with a minimum number of WPs, exhibits remarkable  topological features.
Due to the momentum-dependent nature of light coupling, the positions and band dispersions of the two WPs will evolve with the light amplitude $A_{0}$.
In FM-$z$ MnBi$_2$Te$_4$, the band dispersion of WPs undergoes a light-dependent transition when the intensity $eA_{0}/\hbar$ increases from 0.020 to 0.043 {\AA}$^{-1}$. 
The band dispersions with $eA_{0}/\hbar$ = 0.0225, 0.0300, and 0.0375 {\AA}$^{-1}$ along $\mathrm{\Gamma}$-$\mathrm{Z}$ are illustrated in Figs.~\ref{fig3}(a)-(c), respectively. 
It is found that the dispersion of WP $W^\pm$ transit sequentially from type-II to type-III and finally to type-I. 
To further substantiate this transition, we obtained the Fermi surface in the $k_2=0$ plane of the bulk band structure of FM-$z$ MnBi$_2$Te$_4$ with $eA_{0}/\hbar$ = 0.0225, 0.0300, and 0.0375{\AA}$^{-1}$. 
With energies corresponding to the respective WPs, the isoenergy contours are presented in Figs.~\ref{fig2}(d)-(f). 
In Fig.~\ref{fig2}(d), it is found that when $eA_{0}/\hbar$ = 0.0225{\AA}$^{-1}$, the WPs connect the electron and hole pockets, demonstrating  type II dispersion.
In contrast, as shown in Fig.~\ref{fig2}(f), for $eA_{0}/\hbar$ = 0.0375{\AA}$^{-1}$, the isoenergy contour of this system consists of two isolated points, indicating type-I dispersion of WPs.

\begin{figure}[t]
\includegraphics[width=\linewidth]{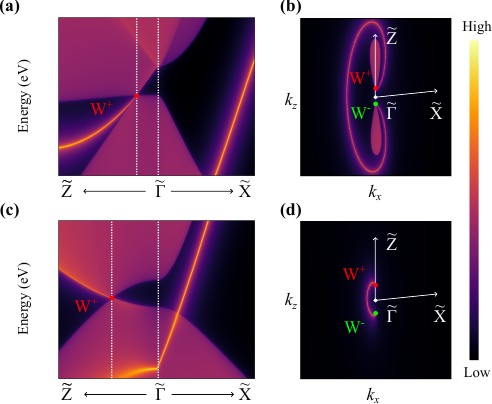}
\caption{(a)-(d) Calculated surface states and corresponding Fermi arcs for the (010) surface of FM-$z$ MnBi$_2$Te$_4$ with irradiation of CPL. 
In plots (a) and (b), $eA_{0}/\hbar$ = 0.0225 {\AA}$^{-1}$ is adopted, while the light intensity is set as $eA_{0}/\hbar$ = 0.0375 {\AA}$^{-1}$ in plots (c) and (d).
}
\label{fig4}
\end{figure}

As the light intensity increases, the characteristic surface states of WPs in the bulk FM-$z$ MnBi$_2$Te$_4$ undergo significant changes. 
The calculated LDOS projected on the semi-infinite (010) surface of FM-$z$ MnBi$_2$Te$_4$ with a light intensity $eA_{0}/\hbar$ of 0.0225 and 0.0375 {\AA}$^{-1}$ are respectively shown in Figs.~\ref{fig3}(a) and \ref{fig3}(c).
It is observed that  upon an increase in light intensity, the slope of the surface state near WP decreases and eventually becomes negative.
Consequently, in contrast to \( eA_0/\hbar = 0.0225 \) Å$^{-1}$ (as detailed in Figs.~\ref{fig4}(b)), when \( eA_0/\hbar = 0.0375 \) Å$^{-1}$, the Fermi arc no longer intersects the high-symmetry line \( \widetilde{\mathrm{Z}} \)-\( \widetilde{\mathrm{\Gamma}} \) but instead directly links the two WPs, as depicted in Figs.~\ref{fig4}(d).
Thus, the length of the Fermi arc is acutely sensitive to variations in light intensity, which affords greater flexibility in manipulating the Weyl semimetal phase in CPL-irradiated FM-$z$ MnBi$_2$Te$_4$.

Finally, in addition to FM-$z$ MnBi$_2$Te$_4$, we have also briefly investigated the band topology of FM-$y$ MnBi$_2$Te$_4$. 
We consider two kinds of CPL, i.e., $\boldsymbol{A}(t)=A_{0}\left[\mathrm{cos}\left(\omega t\right),0,\mathrm{sin}\left(\omega t\right)\right]$ and $\boldsymbol{A}_1(t)=A_{1}\left[\cos(\omega t), \sin(\omega t), 0\right]$. For both cases, we traced the minimal local band gap of FM-$y$ MnBi$_2$Te$_4$.
As shown in Fig.~\ref{fig2}(a), FM-$y$ MnBi$_2$Te$_4$ can be driven to be  semimetal under irradiation of  $\boldsymbol{A}(t)$ or  $\boldsymbol{A_1}(t)$.
For  FM-$y$ MnBi$_2$Te$_4$ irradiated by $\boldsymbol{A}(t)$, when the light intensity $eA_{1}/\hbar$ is between 0.0318 {\AA}$^{-1}$ and 0.0526 {\AA}$^{-1}$, the materials exhibit type-I WSM phase with a pair of WPs.
Similarly, when the light intensity of $eA_{0}/\hbar$ is between 0.0150 {\AA}$^{-1}$ and 0.0240 {\AA}$^{-1}$, FM-$y$ MnBi$_2$Te$_4$ is a Weyl semimetal with one pair of type-I WPs.
In contrast to FM-$z$ MnBi$_2$Te$_4$, the light-induced type-I WPs in FM-$y$ MnBi$_2$Te$_4$ maintain their band dispersion throughout the entire phase diagram.

In the community of condensed matter physics, the  discovery and manipulate of the simplest Weyl semimetal that is experimentally achievable and possesses a single pair of Weyl points is a long-sought goal.
Presently, there are a few proposed candidates, such as HgCr$_2$Se$_4$~\cite{Weyl2}, strained FM MnBi$_2$Te$_4$~\cite{PhysRevLett.122.206401,li2019intrinsic,PhysRevB.103.155118}, and $X$CrTe ($X=$ K, Rb)~\cite{2024arXiv2403}. 
However, due to certain drawbacks, these pioneering candidates face challenges in experimental realization and application.
Here, the realization of the simplest Weyl semimetal in FM MnBi$_2$Te$_4$ under periodic  CPL irradiation take some advantages:
(i) Expect the WPs,  there are no other extraneous
bands crossing the Fermi level.
(ii) The band dispersion, the distance between two WPs, and the length of the Fermi arc are all manipulable by the intensities of the light.
(iii) Inspired by recent experiment where the light-induced anomalous Hall effect in graphene has been observed~\cite{floreal2}, the light-induced and light-manipulate of Weyl semimetal phase would be experimentally achievable.

In conclusion, we theoretically propose that FM MnBi$_2$Te$_4$ exhibits ideal WSM characteristics with  a minimum number of WPs under the irradiation of a CPL. 
Notably, the WPs are controllable by tuning the amplitude of the incident light. 
Specifically, for FM MnBi$_2$Te$_4$ with out-of-plane magnetic orientation, the band dispersion of WPs transitions sequentially from type-II to type-III, and ultimately to type-I, as the light intensity increases. 
Moreover, we show that the profile of the characteristic Fermi arcs connecting the projections of WPs on (010) surface of CPL irradiated FM-$z$ MnBi$_2$Te$_4$ is sensitive to variations in light intensity..
Consequently, the length of the Fermi arc can be effectively controlled by the incident light.
When the magnetic orientation is rotated to the in-plane axis, the FM-$y$ MnBi$_2$Te$_4$ could be driven from FM insulator to Weyl semimetal by CPL irradiating along the $y$ or $z$ axis. 
Different with the case of FM-$z$ MnBi$_2$Te$_4$ where different types of WPs are found, the WPs found in CPL-irradiated FM-$y$ MnBi$_2$Te$_4$ maintains type-I band dispersion until they annihilate..
Our work shows that irradiation of CPL in FM MnBi$_2$Te$_4$ is an efficient means to achieve ideal Weyl semimetals with desired dispersion, which will arouse wide concern of design and light-manipulated non-trivial band topology.

\begin{acknowledgments}
This work was supported by the National Natural Science Foundation of China (NSFC, Grants No. 12204074, No. 12204330, No. 12222402, No. 11974062, No. 12147102, and No. 92365101) and the Natural Science Foundation of Chongqing (Grant No. CSTB2023NSCQJQX0024). D.-S. Ma also acknowledges the funding from the China National Postdoctoral Program for Innovative Talent (Grant No. BX20220367).
\end{acknowledgments}


\bibliography{main}
\end{document}